\documentclass[a4paper,twoside]{article}

\usepackage{epsfig}
\usepackage{subcaption}
\usepackage{calc}
\usepackage{amssymb}
\usepackage{amstext}
\usepackage{amsmath}
\usepackage{amsthm}
\usepackage{multicol}
\usepackage{pslatex}
\usepackage{apalike}
\usepackage{algorithm2e}
\usepackage[bottom]{footmisc}
\usepackage{SCITEPRESS}     
\usepackage{graphicx}
\usepackage{multirow}
\usepackage{titlesec}
\usepackage{balance} 
\usepackage{stfloats}
\usepackage{xcolor}
\usepackage{epsfig}
\usepackage{subcaption}
\usepackage{calc}
\usepackage{amssymb}
\usepackage{amstext}
\usepackage{amsmath}
\usepackage{amsthm}
\usepackage{multicol}
\usepackage{url}
\usepackage{algorithm2e}
\usepackage[bottom]{footmisc}
\usepackage{tabularx}
\usepackage{booktabs}
\usepackage[most]{tcolorbox}

\newtcbtheorem[
  auto counter
]{definition}{Definition}{
  colback=yellow!25,
  colframe=yellow!60,
  fonttitle=\bfseries,
  coltitle=black,
  boxrule=0.5pt,
  arc=2mm,
  left=6pt,
  right=6pt,
  top=6pt,
  bottom=6pt
}{def}

\definecolor{p4blue}{RGB}{147,201,231}    

\begin{document}

\title{Teaching testing seriously in academia}


\author{
\authorname{
Tanja E.J. Vos\sup{1,2}, 
Bart Th. Knaack\sup{1}, 
Beatriz Marín\sup{2}, 
Niels Doorn\sup{1} and 
Nikè van Vugt-Hage\sup{1}
}
\affiliation{\sup{1}Open Universiteit The Netherlands}
\affiliation{\sup{2}Universitat Politècnica de València}
}


\abstract{As systems grow more complex and incorporate AI, testing becomes more critical. Yet testing education in academia remains misaligned with both professional practice and the empirical nature of testing. Current curricula predominantly adopt a rationalist paradigm, emphasizing prescriptive methods and confirmation of expected outcomes. This limits students’ ability to reason critically under uncertainty. 
In this position paper, we argue that testing should instead be taught as an empirical, inquiry-driven professional skill. We propose an instructional design based on the Four-Component Instructional Design (4C/ID) model to support whole-task learning.
We introduce P4TEST, a pedagogical framework that makes explicit the core competencies, epistemic moves, and habits of mind involved in testing, while avoiding prescriptive processes. The paper outlines how P4TEST can guide curriculum design, scaffolding, and assessment in software testing education. }

\onecolumn \maketitle \normalsize \setcounter{footnote}{0} \vfill

\section{Introduction}


Testing is becoming more important in a world where AI-powered
software increasingly shapes critical aspects of society.  
Testing education in academia, however, has largely failed to keep up.
Systematic mappings~\cite{Scatalon2019,garousi2020software,JrA18}, surveys~\cite{MHD2019,SouzaEtAl22,FasolinoEtAl2025} and curriculum studies~\cite{AndradeEtAl19,BarrettETAll23,Tramontanaetal2024} 
consistently report the same patterns: students develop narrow views of testing, lack motivation for good testing, and see the subject as boring. 
Identified causes are that testing is introduced too late in the curriculum, covered only superficially, and systematically misaligned with industrial needs. 

Beyond these issues, however, lies a deeper structural problem: we teach testing using the wrong paradigm, namely rationalism instead of empiricism~\cite{doorn2023towards,vos2023criticaltesters}.
It is our opinion, that the wrong paradigm to teach testing can lead to the majority of the detected problems in testing education. 
In testing, there is rarely one single clear solution. Framing testing as solution-finding, can therefore leave students frustrated and undermine their learning, which in turn negatively affects motivation \cite{Odden21}.



We are not the first to criticise the currently used paradigm. 
The context-driven school~\cite{kaner2001lessons} of testers has been advocating exploratory testing for years. 
Yet, educational materials to teach exploratory testing remain fragmented and are not usable in academic settings. 
Differences between founders and practitioners have further diverged the available perspectives.

One of these trajectories is the Rapid Software Testing (RST) approach. 
After years of blogs and partial materials, an RST book \textit{Taking Testing Seriously}~\cite{BB2025taking} encourages practitioners to think deeply about testing, offering a vocabulary of concepts and philosophical foundation. However, this strength also makes the book unsuitable as a curriculum for beginners. 
RST avoids prescriptive methods and frames testing entirely in terms of heuristics and context-dependent judgement. 
For professional testers, this resonates; for first-year students, with no  experience, it provides too little guidance.
For students it can be difficult to extract the structure they need from the book because it lacks the pedagogical scaffolding that academic education requires: a learning pathway, progression, exercises, and explicit educational structure.
At the same time, there is a pressing need for academic teaching materials that enable serious testing education in bachelor and master programmes, and increasingly -- with the rise of AI -- in primary, secondary and vocational education.


This paper presents a design instrument for testing education that (1) conceptualises testing as a complex whole task, (2) introduces a pedagogical framework (P4TEST) to support curriculum design and (3) operationalises this through the 4C/ID model.


\section{Different paradigms}

In~\cite{ralph2018two} the \textit{rational} and \textit{empirical} paradigms are characterised as disparate constellations of beliefs about how software is and should be created.
We argue that similar constellations exist about 
what testing is and how it should be done. 

Testing education in academia often solely reflects a paradigm aligned with \textit{rationalism} (i.e., emphasising algorithmic problem solving, specification, and methodical application of techniques). Students start with clear requirements, apply predefined procedures, and check expectations. 
This view is reflected in commonly used textbooks on testing \cite{FasolinoEtAl2025}. 
For example, in \cite{Jorgensen2014}, testing is defined as 
\textit{“the act of exercising software with test cases. A test has two distinct goals: to find failures or to demonstrate correct execution”}. 
Subsequently, the book structures testing as a systematic activity applying well-defined techniques for identifying test cases based on specifications or program structure, thereby framing testing as a structured, method-driven activity aimed at evaluating predefined behaviour.
In \cite{AmmannOffutt2018} testing is defined as \textit{the process of applying a few precise, general-purpose criteria to a structure or model of the software}. In the book they organise testing knowledge around coverage criteria over such models. This can be seen as a formalisation of earlier views, such as Beizer’s \cite{Beizer-Techniques} characterisation of testing as \textit{find a graph and cover it}, extending it into a systematic framework of techniques. Other texts, such as \cite{vos2019software}, while placing more emphasis on exploratory testing and acknowledging that the quality of tests depends on the underlying model, still frame testing as a systematic activity: \textit{make a model - pick a coverage criteria - design tests}.

%
These approaches falls short when the goal is to cultivate a critical testing mindset. 
Testing is less about solving problems and more about investigating possible problems. Testing is not about reducing infinite possibilities to one solution (or test suite), but about looking at a solution and diverging into all the things that can possibly go wrong~\cite{rose25drawn}.
Students who learn to verify expected outcomes rarely develop the ability to think about what they might have \emph{not} anticipated. 
Their activity becomes one of confirmation rather than inquiry, and the essence of testing -- investigating, questioning, and making sense of behaviour -- remains underdeveloped.

These shortcomings become more pronounced in the era of AI-based systems. 
With generative AI in particular, the very idea of ``clear requirements'' becomes problematic: the capabilities, boundaries, and failure modes of these systems cannot be fully specified in advance. 
A model may produce biassed results, generate misinformation, or exhibit surprising new abilities. 
Such emergent behaviours raise ethical, technical, and safety concerns that are difficult to foresee. 
In this context, teaching students to merely check expected results is not just insufficient, it prepares them for a kind of system that in the future might no longer exist.

Testing must be seen and taught as a \textit{complex skill}, not as a set of techniques.


\section{Pedagogical tension: scaffolding}
\label{sec:tension} 

Teaching testing as a complex empirical skill means supporting learners in developing \textit{competencies} and specific \textit{testing habits of mind}, all applied in dynamic, context-dependent situations \cite{doorn2023towards}.
Research on learning complex skills ~\cite{vanmerrienboer2024tensteps} shows that effective teaching requires scaffolding to support gradual mastery \cite{vanmerrienboer2024tensteps}.
At first glance, this seems to conflict RST \cite{BB2025taking}, which rejects the idea that testing can be reduced to a linear or prescriptive process. Scaffolding may therefore appear incompatible with testing practice.

This tension, however, exists only at the level of practice, not at the level of education.
Expert behaviour cannot be taught by simply exposing novices to its non-linearity. 
Without guidance, novices struggle to initiate inquiry, notice test-relevant aspects, or make sense of their observations. 
In this context, scaffolding is not a pedagogical whim, but a requirement to enable students to learn.

An illustration of this tension is the discussion of Equivalence Class Partitioning (ECP) and Boundary Value Analysis (BVA). 
Under rationalist framing, they are typically taught as prescriptive methods for deriving tests directly from specifications. 
This frame has contributed to the “unhelpful folklore” in industry 
identified by RST~\cite{BB2025taking}, in which ECP and BVA are used as procedures of reductive steps (that is, identify a class, select a single representative value and declare the domain tested). 
Such an interpretation naturally leads to shallow, confirmation-orientated tests.

However, from an educational paradigm grounded in empiricism, 
ECP and BVA should not be doctrines, but scaffolds. 
Much like introductory programming exercises such as \texttt{hello world} do not represent authentic software development, their educational value 
%
does not lie in reproducing real practice. 
Instead, ECP and BVA provide opportunities for novices to practise simple modelling, explore variation, and confront uncertainty. It should be clear very early on that all models are wrong but some are useful \cite{Box-Models}.
%
The teaching of ECP and BVA should encourage students to consider how faults might arise, why certain distinctions could be meaningful, and how their own modelling decisions shape the behaviour and adequacy of their tests. 








Educational scaffolding supports learning about testing.
The aim is not to convert testing into a stepwise method, but to support students in playing the epistemic game~\cite{collins1993epistemic} of testing: 
gradually acquiring and coordinating the moves needed for effective testing. By understanding testing as an epistemic game, we can better assess when students genuinely engage in testing rather than merely reproducing answers.
It is important to emphasise that, while the described pedagogical tension can be addressed at the level of education through carefully designed scaffolding, it cannot be eliminated. Any scaffold introduced to support learning carries the risk of being mistaken for a prescription, particularly when learners and instructors seek certainty in domains characterised by uncertainty. 


To carefully design scaffolding, we use 4C/ID, the Four Component Instructional Design framework~\cite{vanmerrienboer2024tensteps}.

\section{4C/ID}
\label{section.4cid}

4C/ID is a framework developed for the design of educational programmes aimed at learning complex  skills~\cite{vanmerrienboer2024tensteps}. 
In contrast to approaches that organise curricula around isolated objectives or techniques, 4C/ID takes integrated professional performance as its point of departure.

Central to 4C/ID is the notion of a \textit{whole task} as an integrated professional activity.
One of the 4 components are the \textit{learning tasks}, which are educational instantiations of whole tasks. 
These tasks form the backbone of the curriculum and are sequenced in classes of increasing complexity, variability, and decreasing scaffolding.

To support learning of the whole task, 4C/ID moreover distinguishes the following components:
%
\textit{Supportive information}, the resources that are referred to as non-recurrent aspects of task performance, reflecting their non-routine and context-dependent nature.
\textit{Procedural information}, the guidance for carrying out recurrent actions during task performance, enabling learners to act without having to reconstruct these actions each time (think about memorizing the multiplication tables).
\textit{Part-task practice}, the drills that may be used selectively to support fluency in specific aspects of task performance.

Together, these components provide a structure for designing instructional support for complex skills without decomposing that performance into linear procedures. While the model is composed of these 4 components, their emphasis depends on the domain and the nature of the tasks involved.

In the context of testing, learning primarily concerns non-routine, context-sensitive activities that require judgement, sensemaking, reflection and adaptation. As a result, whole-task learning and supportive material are central, whereas procedural material and part-task practice is used only sparsely. 




\begin{figure*}[ht]
    \centering
    \includegraphics[width=\linewidth]{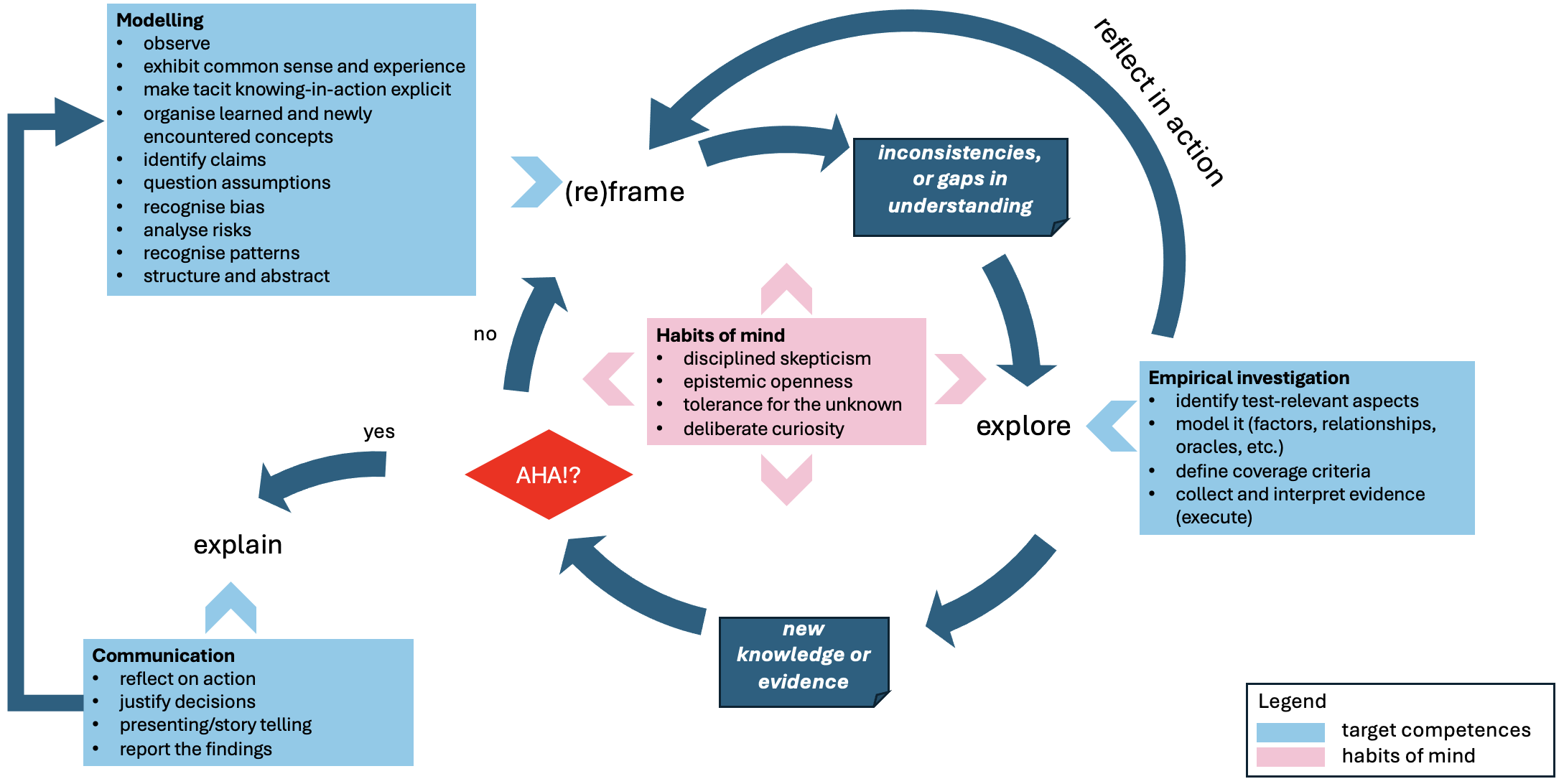}
    \caption{P4TEST - Pedagogical framework for testing}
    \label{fig:P4TEST}
\end{figure*}

\section{Testing as a whole task}

4C/ID starts with the following question to define the whole task:
\noindent \textit{What must someone be able to do in the real world?} 

Existing definitions of testing in standards and in practitioner literature  describe testing in terms of goals, processes, or philosophies. 
Although these definitions can be valuable, they are insufficient for instructional purposes. 
%
%
We adopt the following:

\begin{definition}{Testing as a whole task}{testing}
\textit{To produce and effectively communicate, within limited time, an evidence-based evaluation of the quality and risks of software in complex and uncertain contexts through modelling, empirical investigation, sensemaking, and reflection-in-action.}
\end{definition}


This definition aligns with practitioner perspectives on testing \cite{BB2025taking} where testing is seen as \textit{the process of evaluating a product by learning about it}. The components of the definition make this learning process explicit in terms of the activities (competences) that constitute testing practice. 
\textit{Modelling} is the practice of reasoning about the world using abstraction and structure.
A model leaves out details that are not necessary for a specific goal and is therefore a simplification of reality. 
%
\textit{Empirical investigation} is the collection of evidence through sensory experience (e.g., observation of and interaction with reality), guided by questions and models \cite{MendezAKA24}.
\textit{Sensemaking} is the dynamic process of framing, exploring, and revising explanations in order to “figure something out” and resolve gaps in one’s understanding \cite{odden2019defining}.
\textit{Reflection-in-action} refers to the spontaneous, “on-the-fly” re-framing of action while engaged in practice, in response to the situation as it unfolds \cite{Schon87}.

If testing is understood as a form of learning, it is not surprising that patterns described in general learning theories can also be recognised in testing. For example, Kolb’s experiential learning theory \cite{Kolb_Kolb_2022} characterises learning as an iterative process involving experience, reflection, and conceptualisation.
In the next section we will build on this insight by making explicit what learning looks like in the specific context of testing, by operationalising the whole-task definition into
a pedagogical framework (P4TEST) for
for instructional design through 4C/ID.


%
%
 
\section{P4TEST}
\label{sec:P4TEST}

P4TEST (Figure \ref{fig:P4TEST}) is a pedagogical framework that structures our vision for testing education. The framework provides a rationale for curriculum design and helps students understand that design.
We want to emphasise once more that P4TEST is a learning scaffold: it scaffolds thinking about testing,  
\textit{not} testing itself. Rather, students must learn to both use it and outgrow it.

P4TEST contains all the elements of the whole-task testing definition (Def. \ref{def:testing}):
the \textit{modelling} and \textit{empirical investigation} elements of Def. \ref{def:testing} are represented as core competencies, while \textit{sensemaking}
and \textit{reflection-in-action} are captured through the iterative dynamics of the actions that connect them. 
Finally \textit{communication}, as a competency, consolidates the resulting insights into explaining and forming an evaluative judgement. 

Students begin by framing and modelling what is already known, drawing on observations, common sense, prior learned concepts, and other competencies that are in the modelling box in Figure \ref{fig:P4TEST}. 
%
In doing so, they inevitably encounter gaps, inconsistencies, or anomalies that require understanding through exploration. 
During exploration, learners investigate these gaps by designing experiments, to gather evidence. 
This process is inherently dynamic: through reflection-in-action, learners may need to reframe their initial understanding, adjust their models, and iterate on their experiments. 
This process will iterate until an AHA! moment of insight~\cite{kounios2009aha} after which they will consolidate the results into explanation, through communication and reflection-on-action.

In P4TEST, modelling occurs at different levels of scope and purpose. As a core competency, modelling concerns the tester’s understanding of the overall testing situation and what we already know or have discovered about the SUT, i.e. its context of use, relevant risks, quality concerns, testing goals, etc. During empirical investigation, modelling takes a more focused form, supporting the exploration of specific test-relevant aspects, such as particular behaviour, pattern, data domain, condition, or assumption of interest. This distinction is consistent with the way RST \cite{BB2025taking} differentiates between reasoning about the system as a whole (\textit{product space}) and more focused reasoning used to support specific investigative questions (\textit{assessment space}).

At the centre of Figure \ref{fig:P4TEST} are the \textit{testing habits of mind} that we have identified as characteristic of critical testers~\cite{doorn2023towards}: 
\textbf{disciplined scepticism} the practice of questioning convincing results making tacit assumptions explicit, actively challenging evidence, and turning doubt into a disciplined engine of inquiry;
\textbf{epistemic openness} the willingness to revise conclusions when new evidence appears;
\textbf{tolerance for the unknown} the capacity to sustain inquiry and productivity when rules are incomplete, ambiguous, or misleading. Instead of being paralysed by uncertainty, learners use it as fuel for more testing;
\textbf{deliberate curiosity} the intentional drive to notice anomalies, to pose ``what if?'' questions, to explore alternatives beyond the obvious, and to define experiments.



\section{From whole task to task classes}
\label{section:from-p4test-to-id}

\begin{table*}[ht]
\centering
\small
\caption{Task classes in the 4C/ID-based testing curriculum}
\label{tab:task_classes}
\begin{tabularx}{\textwidth}{p{0.65cm}p{2.4cm}p{2.2cm}XX}
\toprule
\textbf{TC} &
\textbf{Nature} &
\textbf{Type} &
\textbf{Goal} &
\textbf{Scaffolding} \\
\midrule
TC1 &
Analysis of worked-out testing examples. &
Whole-task analysis (evaluative) &
Recognise P4TEST elements in testing practice and identify alternative test ideas &
Very high; explicit prompts, guiding questions, visible P4TEST \\
\addlinespace
TC2 &
Guided testing of SUTs &
Whole-task execution (guided) &
Integrate recognised P4TEST elements through basic modelling, exploration, and reporting &
High; structured tasks, frequent feedback, explicit support, through (reflective) questions. \\
\addlinespace
TC3 &
Testing of more complex and less constrained systems &
Whole-task execution (increasingly autonomous) &
Develop testing strategies, make informed trade-offs, and engage in reflection-in-action during empirical testing &
Low; selective conceptual support, minimal prompts \\
\addlinespace
TC4 &
Authentic testing in a realistic context &
Whole-task execution (autonomous) &
Produce, communicate, and justify an evidence-based evaluation of software quality and risk &
Minimal; performance criteria only \\
\bottomrule
\end{tabularx}
\end{table*}

4C/ID is organised around \textit{learning tasks} (LT), which are educational instantiations of the whole task.
They are grouped into \textit{Task Classes} (TC) and sequenced according to increasing complexity and variability, while support is systematically faded.
%
We have designed the four TCs in Table \ref{tab:task_classes}. 
These TCs make the curriculum gradually increase in complexity, autonomy, and responsibility. 

\paragraph{TC1 - Analysis of worked-out examples}

Students are introduced to testing as an empirical activity through the analysis of \textit{worked-out examples} that illustrate elements of P4TEST. Rather than performing tests, students examine the work of an experienced tester. 
The worked-out examples range from everyday artefacts (e.g., a pen or a handbag, see Figure \ref{fig:TC1.pen}), to pseudo-code representations (see Figure \ref{fig:TC1.naur} for an example), to small executable software systems such as a simple to-do application.
Instructional support in this task class is intentionally strong. The examples are accompanied by explicit commentary, guiding questions, and reflective prompts that draw attention to critical moves in testing, such as 
the emergence of test ideas, 
the identification of assumptions, 
the recognition of gaps in understanding, etc.
%
The worked-out examples deliberately foreground (and gradually introduce) conceptual resources, starting with the concept of quality, mind maps, product coverage outlines (PCOs), simple test models like decision tables, combinatorial theory and heuristics.
The emphasis is not on applying techniques mechanically, but on developing the capacity to select and adapt modelling approaches and heuristics in response to the characteristics of the SUT.


\begin{figure}
\small
\begin{tcolorbox}[title=Example TC1 task - testing a pen,
colback=gray!5,
colframe=p4blue,
boxrule=0.5pt]

\textbf{The learning (whole) task -} Students analyse a worked example in which an experienced tester investigates and communicates about the behaviour of a pen (but can be any everyday artefact). They trace how the tester moves back and forth within P4TEST during testing. After reading, the students have to test a relevant aspect themselves and write down their steps.

\textbf{A testing story -}
The tester starts writing on a paper and notices the pen briefly hesitates before producing ink. This raises the question whether such hesitation is always a quality issue. The tester reflects on the purpose of the pen (e.g.\ writing, drawing, promotional use) and narrows the focus to writing, where ink flow is central.
Further questions emerge as gaps in understanding: does the hesitation only occur at the start? Does the writing surface matter (paper, cardboard, glossy, dirty, clean)? Do angle or pressure influence ink flow? Does the state of the pen (new, cold, unused) matter? These questions are incrementally organised in a mind map that serves as a model for further testing.
The tester then focuses on one test-relevant aspect:
\textit{Does the pen work on clean paper under different writing conditions?}
To explore this, the tester considers typical writing behaviour and varies combinations of speed, pressure, and angle.
Under high pressure and a steep angle, ink blobs appear, while under normal conditions writing remains consistent. The tester revises the model to include both insufficient and excessive ink flow as relevant quality concerns.

\textbf{Supportive information about - } P4TEST, mindmaps, the use of heuristics, the concept of quality and basic combinatorial theory.
\end{tcolorbox}
    \caption{Example of a TC1 task}
    \label{fig:TC1.pen}
\end{figure}


\begin{figure}
\small
\begin{tcolorbox}[title=Example TC1 task - testing an algorithm,
colback=gray!5,
colframe=p4blue,
boxrule=0.5pt]

\textbf{The learning (whole) task -}  
Students analyse two worked examples of testing Naur’s text formatting algorithm. 
The algorithm \cite{Naur1969} determines its behaviour based on several conditions 
(e.g., input, system state, and configuration). Although it appears straightforward, 
it contains subtle errors and has long been used to illustrate different testing approaches \cite{GoodenoughGerhart1975}. 

Students compare the examples to analyse how different approaches move through P4TEST and how understanding 
of the system emerges through modelling, exploration, and revision. 

\textbf{The first testing story -}  
An exploratory tester begins by implementing the algorithm and performing simple 
experiments, such as varying the viewport size. By observing patterns in the output, 
several issues become visible, including unexpected leading spaces, incorrect handling 
of long words, and inconsistent treatment of line breaks. Test ideas emerge from 
these observations, guided by tacit knowledge, heuristics, and incremental exploration.

\textbf{The second testing story -}  
In a second example, the tester constructs a decision table to model the behaviour 
of the algorithm. Conditions such as buffer size, line length, and input characters 
are analysed systematically. Through this modelling process, implicit assumptions 
and gaps in the specification become visible, leading to new test ideas (e.g.\ handling 
of consecutive blanks, words of maximum length, and end-of-file conditions).

\textbf{Supportive information about - } the heuristic SFDIPOT\footnote{a heuristic originating from RST used to stimulate test ideas by considering Structure, Function, Data, Interfaces, Platform, Operations, and Time.}, PCOs, decision tables, oracles.
\end{tcolorbox}
    \caption{Example of a TC1 task}
    \label{fig:TC1.naur}
\end{figure}

\begin{figure}
\small
\begin{tcolorbox}[
title=Example TC2 task -- a random number generator,
colback=gray!5,
colframe=p4blue,
boxrule=0.5pt,
arc=2mm]

\textbf{The learning (whole) task -}  
Students test a web-based random number generator recognizing the P4TEST activities, and learn that
a system that initially appears simple can reveal multiple layers of complexity.

\textbf{The SUT -}  
At first glance, the system appears simple: users specify a range and generate numbers. 
However, guided by reflective questions, students are encouraged to investigate the system more deeply, 
revealing multiple layers of complexity. 

They uncover aspects such as properties of randomness (e.g.\ uniformity, predictability of seeds), 
decimal precision and rounding behaviour, and edge cases (e.g.\ very small or extreme ranges). 
They also consider appropriate oracles for evaluating correctness.
In addition, the system provides multiple entry points for input (GUI and URL), 
raising questions about consistency between different interaction paths.

\textbf{Supportive information about - } other heuristics like CRUD\footnote{data-related behaviour by considering Create, Read, Update, and Delete operations} and HICCUPS\footnote{Oracles related to History, Image, Claims, Comparable products, User Desires, Product, Purpose, and Statutes, from \cite{BB2025taking}}, ECP and BVA
\end{tcolorbox}
    \caption{Example of a TC2 task}
    \label{fig:random}
\end{figure}

\paragraph{TC2 - Guided testing of simple SUTs}
that are intentionally limited in scope and complexity to allow students to focus on testing, rather than on technical or infrastructural complexity. 
Students are supported with conceptual resources for constructing and refining test models, such as input domain models (e.g. ECP, BVA) and other heuristics. 
These are used not only to generate test ideas, but also to structure reasoning about the system, make assumptions explicit, and explore how different interpretations of behaviour may influence testing.
Following \cite{BB2025taking}, and the discussion in Section \ref{sec:tension}, it is emphasised that the models on which techniques such as ECP and BVA are based are inherently hypothetical. 
As a result, coverage reflects the model that was constructed, rather than guaranteeing adequate testing of the SUT.
Moreover, support for documenting findings and communicating results in a clear and meaningful way will be provided.
Instructional support remains substantial but shifts in character. 


\paragraph{TC3 – Testing of more complex and less constrained SUTs}

Students engage with SUTs of increased complexity and reduced structural guidance. 
They will be supported with resources on more expressive test models, such as classification trees, state-based representations and graph-based models (especially event-based like the tools used in \cite{doorn2023towards}). Students are encouraged to combine all the models they have learned to reason about test coverage, anticipate system behaviour, and explore the implications of different modelling choices. The emphasis lies on managing increasing complexity.
Instructional support is further reduced in this phase.

\begin{figure}
\small
\begin{tcolorbox}[title=Example TC3:  testing more complex systems,
colback=gray!5,
colframe=p4blue,
boxrule=0.5pt]

\textbf{The learning (whole) task -}  Students test a small web application that
simulates a point-of-sale system calculating the change returned
after a payment.

\textbf{The SUT -}
The system involves several interacting aspects such as payment
type, currency, rounding behaviour, and the availability of
denominations in the cash drawer. These means 
different models can be used to investigate the system, including
input domains, parameter combinations, and simple flow models.
The goal of the assignment is to practise integrating modelling and
empirical investigation while testing a real system.

\includegraphics[width=\linewidth]{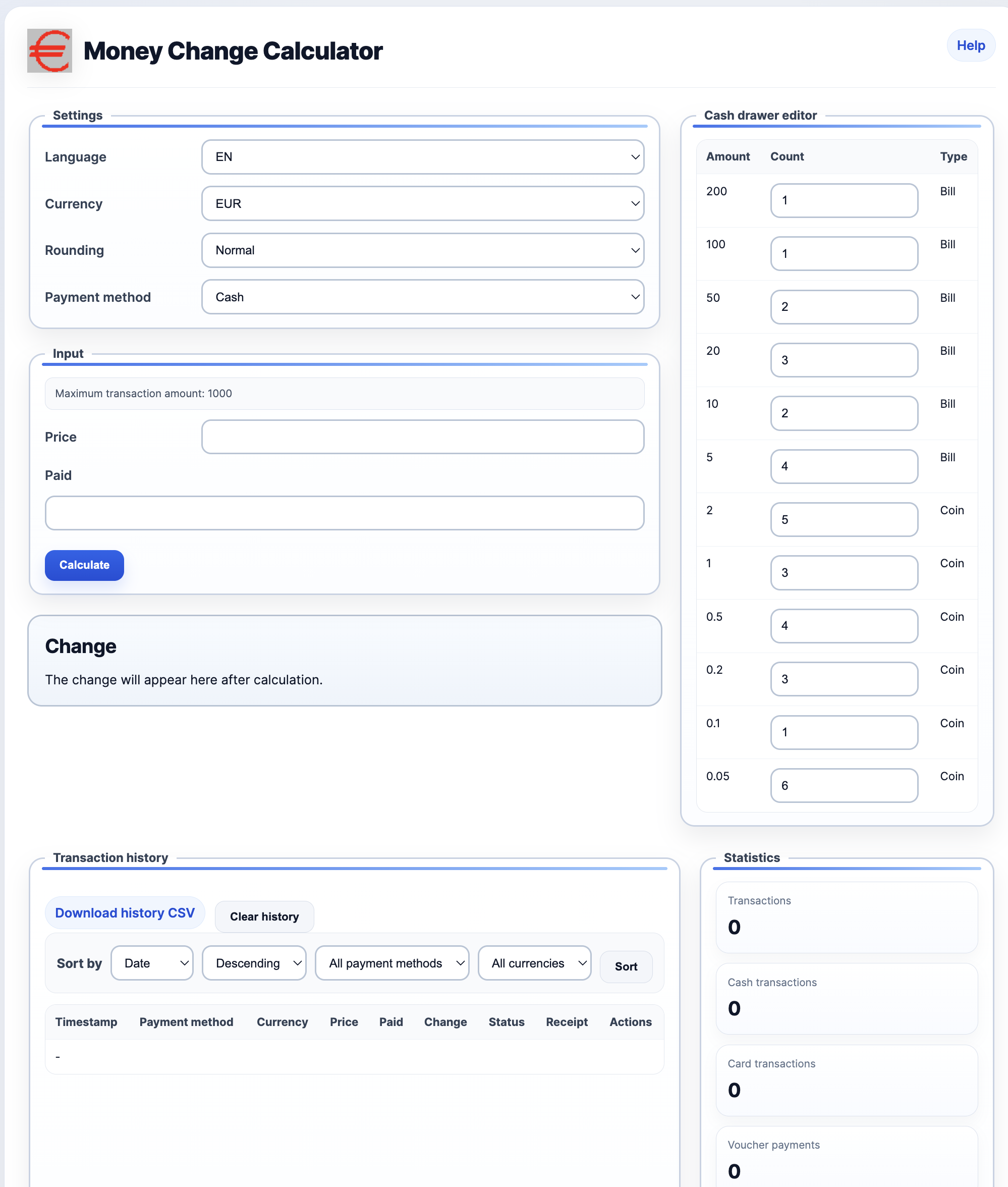}

\textbf{Supportive information about - } Classifcation trees, event-flow

\end{tcolorbox}
    \caption{TC3 example (MoneyChange)}
    \label{fig:moneychange}
\end{figure}

\paragraph{TC4 – Authentic testing in a realistic context}
Students engage in an authentic testing assignment that closely resembles professional testing practice. 
At this stage, students operate with a high degree of autonomy. They are responsible for identifying relevant testing goals, selecting and applying appropriate models, and making informed decisions about what to test, how to test it, and how to interpret results. Students must justify their choices and adapt their approach in response to emerging findings and contextual constraints.
Instructional support in this class, depending on the level (bachelor or master), can be extended towards more expressive test models from academia, such as Labelled Transition Systems (LTS) as models and (u)ioco ((universal) input-output-conformance) as implementation relation \cite{Tretmans2017}.

%


\section{Supportive didactics}
Across the tasks, students engage with the whole task, while gradually developing the competencies represented in the light blue boxes of P4TEST. 
The associated testing habits of mind (Figure~\ref{fig:P4TEST}) are implicitly exercised, but experience shows that they require explicit didactic reinforcement. 
These habits are systematically underdeveloped in traditional curricula.

\emph{Disciplined scepticism} is often discouraged. 
Existing curricula tend to eliminate doubt by confirming assumptions or treating them as unquestionable background knowledge. 
From an empiricist perspective, this logic should be inverted: doubt should become the engine of inquiry, and progress should be measured by uncovering hidden assumptions.
\emph{Epistemic openness} is needed such that inquiry does not collapse into endless scepticism, but includes a readiness to revise conclusions when evidence is sufficient. 
\emph{Tolerance for the unknown} is rarely addressed in education, despite its importance in professional practice and its recognition in other domains such as medicine~\cite{Pa22}. 
Educational settings typically reduce uncertainty by providing complete specifications, unambiguous tasks, and tightly framed assignments. 
In contrast, contemporary software systems -- particularly AI-based systems -- are characterised by incomplete data, unclear requirements, and unstable behaviour. 
Students must therefore learn to remain productive under uncertainty rather than expecting it to be resolved for them.
Although \emph{curiosity} is widely praised as an educational virtue~\cite{Na22}, it is seldom cultivated or assessed systematically. 
Students are commonly rewarded for producing correct answers, after which inquiry stops. 
Testing, however, requires curiosity: the ability to continue asking questions even when an answer appears satisfactory.


To explicitly cultivate these testing habits of mind, we propose puzzle-based learning as a supportive didactic strategy. 
Riddles, brain teasers, and other puzzles have a long tradition in education~\cite{MM08}. 
While game-based learning approaches for testing (e.g., CodeDefenders~\cite{Fra19}, VU-BugZoo~\cite{Si25}) have successfully increased engagement, they primarily emphasise on bug finding and competition.
Puzzle-based learning differs in that it confronts learners with situations that cannot be solved through routine methods. 
Such activities have demonstrated cognitive benefits, including enhanced attention, memory, and cognitive flexibility~\cite{RK12}. 
In prior work~\cite{DVMD25}, we describe a structured set of puzzle-based activities for testing education, designed to be reusable and pedagogically explicit, including learning goals, preparation, mechanics, and debriefing.
Our puzzles share three key characteristics:
(1) They admit multiple plausible solutions, challenging students to question assumptions and explore alternative explanations rather than settling on the first answer.
(2) They create multiple moments of insight (\emph{aha!}'s), which improve engagement and support learning~\cite{kounios2009aha}.
(3) They are easy to remember, increasing the likelihood that the associated concepts and habits are retained~\cite{MM08}.

\section{Assessing testing as a complex skill}

In line with 4C/ID, assessment should focus on whole-task performance rather than on isolated techniques or procedures.

Traditional written exams are not well suited for assessing testing competency \cite{French+24}.
Exams mainly measure declarative knowledge and factual recall, such as definitions and techniques.
They do not capture how students reason while testing, how they make decisions, how they adapt their approach when new evidence emerges, or whether they have developed the required habits of mind.

Written test reports are better suited to represent the whole task, but also have limitations.
Reports describe testing after the fact and therefore hide reasoning-in-action.
Students may rationalise decisions retrospectively, while epistemic moves and activation of habits of mind remain invisible \cite{Rusman2023}.
With the increasing use of generative AI, it is also difficult to determine whether a report reflects the student’s own testing activity.


To address these limitations, we propose \textit{video-based assessment}.
In medical education, these are widely used to support the evaluation of clinical reasoning: the process by which healthcare professionals collect and interpret information, deal with uncertainty and incomplete patient information, generate and test hypotheses, and make context-sensitive decisions about patient care \cite{Robbrecht25}.
These characteristics of clinical reasoning mirror empirical testing.
Adopting this assessment form, puts focus on observable epistemic moves during testing rather than on test artefacts or defect counts.


To make moves visible, students will record testing sessions, including screen activity and think-aloud verbalisation, capturing testing as it unfolds in practice. This enables students’ testing to be observed and assessed as whole-task performance.

In line with P4TEST, assessment criteria should attend to dimensions such as framing and modelling decisions, handling of uncertainty, questioning assumptions, evidence generation and interpretation, reflection-in-action (e.g. reframing or course correction), and communication of an evaluative judgement.

Assessment is organised through structured peer evaluation. Students review each other’s recorded testing sessions with instructor oversight to support calibration, consistency, and reliability.

This assessment approach aligns with whole-task performance assessment in 4C/ID, reduces reliance on AI-sensitive written artefacts, and reinforces learning by engaging students in authentic testing practice.

\section{Conclusions and future work}

This paper presents a design instrument, not a prescriptive teaching method.
By defining testing as a whole task and using 4C/ID for instructional design, this paper provides a basis that can support educators with redesigning their testing curricula.


P4TEST supports this redesign by indicating which forms of modelling and inquiry should be foregrounded at different stages of learning. In early task classes, modelling choices may be constrained and made explicit, while later task classes require students to construct, adapt, and justify their own models.
Importantly, P4TEST does not prescribe a sequence of testing activities; instead, it helps educators reason about which cognitive and epistemic activities a learning task should elicit, while preserving the empirical and context-dependent nature of testing practice.



As future work, behavioural logging during the execution of testing activities will be explored as a complementary source of performance evidence, alongside video-based assessment. Moreover, experiments are planned to evaluate the usefulness of P4TEST in bachelor and master programmes.

\section*{\uppercase{Acknowledgements}}

The authors would like to thank James Bach for his valuable comments and thoughtful feedback on earlier versions of this paper. 
His insights helped improve the clarity and overall quality of the work.

\bibliographystyle{apalike}
{\small
\bibliography{example,IB3202,puzzle}}

\end{document}